# Robustness Evaluation of a Foundation Segmentation Model Under Simulated Domain Shifts in Abdominal CT: Implications for Health Digital Twin Deployment

Sanghati Basu
Healthcare Informatics, University of Illinois Springfield
https://github.com/SANGHATI23/sam-brats-robustness-audit

**Abstract**

Foundation segmentation models such as the Segment Anything Model (SAM) have demonstrated strong generalization across natural images; however, their robustness under clinically realistic medical imaging domain shifts remains insufficiently quantified. We present a systematic slice-level robustness audit of SAM (ViT-B) for spleen segmentation in abdominal CT using 1,051 nonempty slices from 41 volumes in the Medical Segmentation Decathlon. A standardized ground-truth-derived bounding-box protocol was used to isolate encoder robustness from prompt uncertainty. Controlled perturbations simulating inter-scanner variability, including Gaussian noise, blur, contrast scaling, gamma correction, and resolution mismatch, were applied across ten conditions. The clean baseline achieved a mean Dice score of 0.9145 (95% CI: [0.909, 0.919]) with a failure rate of 0.67%. Across all perturbations, the absolute mean ΔDice remained below 0.01. Paired Wilcoxon signed-rank tests with Benjamini-Hochberg false discovery rate correction identified statistically significant but small-magnitude changes under selected conditions, while McNemar analysis showed no significant increase in failure probability. These findings indicate that SAM exhibits stable segmentation behavior under moderate CT domain shifts, supporting its role as a robust foundation baseline for medical image segmentation research. As health digital twins increasingly incorporate foundation segmentation models for anatomical modeling and organ-level monitoring, formal characterization of robustness under real-world imaging variability is a necessary step toward trustworthy deployment.

## 1. Introduction

Foundation models pre-trained on large-scale data have emerged as general-purpose tools for vision tasks, with the Segment Anything Model (SAM) [1] representing a prominent example. SAM has been applied to medical image segmentation with encouraging results [2]; however, clinical deployment demands rigorous characterization of model robustness under distribution shift, a known failure mode of deep learning systems [3], [4].

Domain shifts in clinical CT arise from differences in scanner hardware, acquisition protocols, reconstruction kernels, and patient positioning. Prior work on robustness in medical imaging has focused primarily on fully supervised models [5]; systematic evaluation of foundation model stability under such shifts has not been thoroughly addressed.

Health digital twins, dynamic computational representations of individual patients that evolve over time and integrate multimodal data streams, increasingly rely on medical image segmentation as a core anatomical modeling component [7]. For a digital twin to maintain a valid, continuously updated representation of organ structure, the segmentation models it depends upon must demonstrate stable performance across the heterogeneous imaging conditions encountered in real-world clinical deployment. Inter-scanner variability, arising from differences in hardware, acquisition protocols, and reconstruction parameters, represents one of the most prevalent sources of domain shift that twin systems must contend with. Establishing that foundation segmentation models such as SAM can withstand such variability is therefore not merely a model evaluation question, but a prerequisite validation step for trustworthy digital twin infrastructure.

This paper makes the following contributions: (i) a standardized GT-derived box-prompt evaluation protocol that isolates encoder robustness independent of prompt uncertainty; (ii) a large-scale slice-level robustness audit across 10 controlled perturbation conditions; (iii) quantified ΔDice with paired statistical testing (Wilcoxon signed-rank + McNemar with FDR correction); and (iv) a cumulative reliability profile characterizing clinical stability thresholds.

## 2. Methods

### 2.1. Dataset

Experiments were conducted on Task 09 (Spleen) from the Medical Segmentation Decathlon [6]. The dataset contains 41 abdominal CT volumes with corresponding binary spleen segmentation masks. A total of 1,051 non-empty axial slices (containing at least one foreground voxel) were extracted for evaluation. Dataset characteristics are summarized in TABLE I.

**TABLE I. DATASET CHARACTERISTICS AND PREPROCESSING PARAMETERS**

| Attribute | Value |
| --- | --- |
| Imaging modality | Computed Tomography (CT) |
| Dataset source | Medical Segmentation Decathlon (Task09-Spleen) |
| Total volumes (cases) | 41 |
| Non-empty slices evaluated | 1,051 |
| Slice resolution | 512 × 512 pixels |
| HU windowing | Level = 50 HU, Width = 400 HU |
| SAM model variant | ViT-B (sam_vit_b_01ec64.pth) |
| Prompt type | GT-derived bounding box – single mask output |
| Prompt rationale | Isolates encoder robustness independent of prompt uncertainty |

*HU = Hounsfield Unit; GT = ground truth; ViT = Vision Transformer; SAM = Segment Anything Model.*

## 2.2. Preprocessing and SAM Inference

CT intensities were windowed to a level of 50 HU and width of 400 HU, clipped to [0, 255], and converted to 3-channel RGB format by replicating the grayscale channel. The SAM ViT-B checkpoint was used with a single bounding box prompt derived from the ground-truth mask extents. This controlled box-prompt protocol was selected to isolate encoder robustness independent of prompt uncertainty, providing a clean upper-bound estimate of model stability.

## 2.3. Robustness Protocol

Robustness was quantified using the slice-level metric ΔDice = Dice_perturbed − Dice_clean, computed for each perturbation condition against the clean baseline. Ten conditions were evaluated across five perturbation types designed to simulate realistic inter-scanner variability. Perturbation parameters are detailed in TABLE II.

**TABLE II. DOMAIN SHIFT PERTURBATION PROTOCOL AND PARAMETER SETTINGS**

| Perturbation | Clinical Analog | Parameter | Levels | Severity |
| --- | --- | --- | --- | --- |
| Gaussian Blur | Scanner PSF variation | Kernel size k | k = 3, 7 | Low/Moderate |
| Additive Gaussian Noise | Detector noise | Std. dev. s | s = 10, 25 | Low/Moderate |
| Downsample + Upsample | Resolution mismatch | Scale factor | 0.5x, 0.25x | Low/Moderate |
| Contrast Scaling | Intensity normalization | Multiplier a | a = 0.8, 1.2 | Low |
| Gamma Correction | Brightness non-linearity | Exponent g | g = 0.8, 1.2 | Low |

*PSF = point spread function; HU = Hounsfield Unit. Two severity levels per perturbation type.*

## 2.4. Evaluation Metrics

Segmentation quality was assessed using the Dice Similarity Coefficient (DSC) and Intersection over Union (IoU):

$DSC = 2|X \cap Y| / (|X| + |Y|)$   (1)
$IoU = |X \cap Y| / |X \cup Y|$   (2)

Slice-level failure was defined as Dice < 0.5. A representative segmentation result is illustrated in Figure 1.

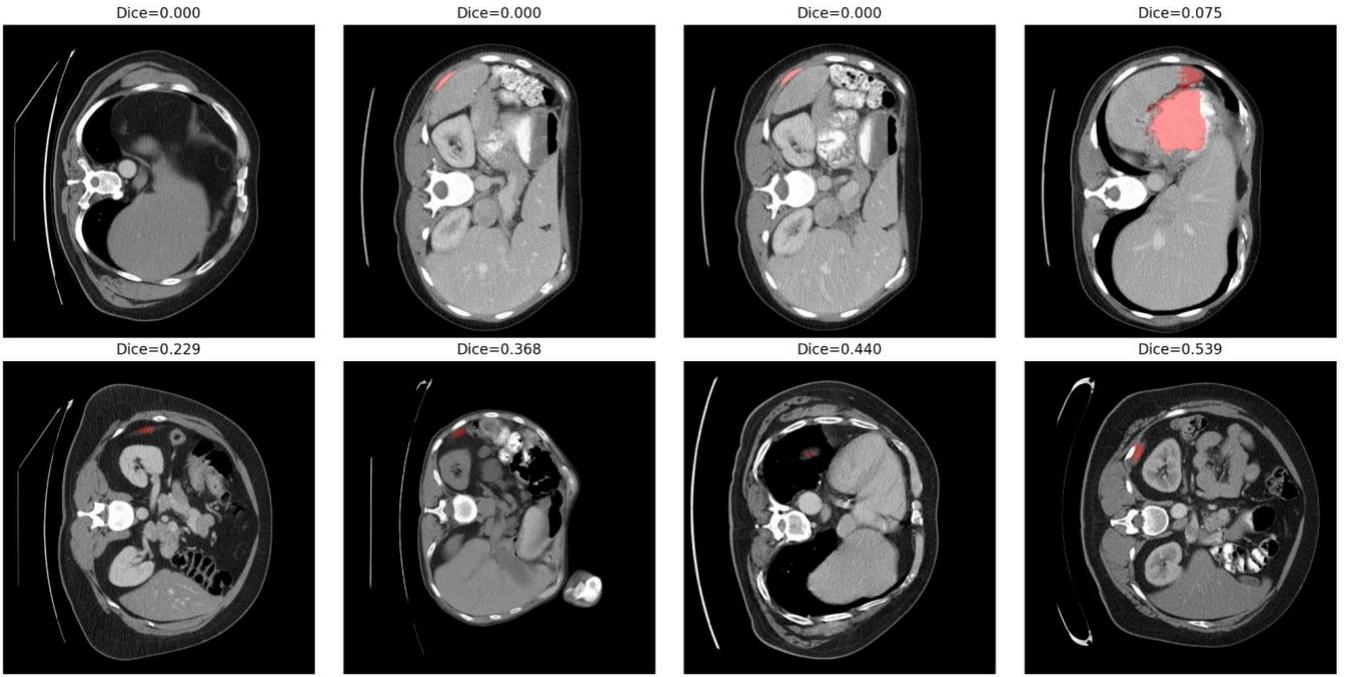

*Fig. 1. Worst-case failure slices (Dice < 0.55) from the clean baseline evaluation. Top row (left to right): Dice = 0.000, 0.000, 0.000, 0.075. Bottom row: Dice = 0.229, 0.368, 0.440, 0.539. Red overlay indicates SAM prediction where present. Failures cluster in anatomically ambiguous slices with minimal or absent spleen cross-section, confirming that collapse cases are boundary-slice artifacts rather than systematic segmentation errors.*

### 2.5. Statistical Analysis

Bootstrap resampling (10,000 iterations, slice-level) was used to estimate 95% confidence intervals for mean Dice and IoU. For each perturbation condition, a paired Wilcoxon signed-rank test was applied to slice-level ΔDice values, testing whether the median change differed from zero. Changes in failure status (Dice < 0.5) were evaluated using McNemar's test for paired binary outcomes. Both sets of p-values were independently corrected for multiple comparisons using the Benjamini-Hochberg false discovery rate (FDR) procedure across the 10 perturbation conditions. Effect sizes were quantified using rank-biserial correlation (r). Code is available from the corresponding author upon request.

## 3. Results

### 3.1. Baseline Segmentation Performance

Across 1,051 non-empty spleen CT slices derived from 41 volumetric scans, SAM (ViT-B) achieved a mean Dice score of 0.9145 (95% bootstrap CI: [0.909, 0.919]) and mean IoU of 0.8532 (95% CI: [0.846, 0.860]). The median Dice score was 0.9436 (IQR: 0.9117-0.9621), confirming that most slices demonstrated high spatial overlap with the ground truth. The slice-level failure rate (Dice < 0.5) was 0.67% (7/1,051; 95% CI: [0.27%, 1.38%]). 81.1% of slices (852/1,051) achieved Dice ≥ 0.9, while 99.3% exceeded Dice ≥ 0.5. These results establish a strong baseline performance prior to robustness analysis.

### 3.2. Robustness Under Controlled Domain Shifts

Performance under simulated domain shifts is summarized in TABLE III and visualized in Figure 2. Slice-level ΔDice values were evaluated using paired Wilcoxon signed-rank tests with Benjamini-Hochberg FDR correction across the 10 perturbation conditions.

Mild Gaussian blur (k=7) and moderate downsampling (×0.5) demonstrated small but statistically significant improvements relative to the clean baseline (median ΔDice > 0, FDR-adjusted $p < 0.05$). The observed improvement under blur likely reflects suppression of high-frequency noise artifacts in the preprocessed CT slices rather than enhanced semantic understanding. In contrast, gamma reduction (γ=0.8) and contrast increase (×1.2) produced modest but statistically significant degradation (median ΔDice < 0, FDR-adjusted $p < 0.05$). Effect sizes were consistently small (rank-biserial correlation |r| range: 0.02-0.08), and the absolute magnitude of mean ΔDice remained below 0.01 across all conditions, indicating limited practical impact on segmentation accuracy.

TABLE III. SEGMENTATION PERFORMANCE UNDER CONTROLLED DOMAIN SHIFTS (N = 1,051 SLICES)

| Condition | Type | Mean Dice | ΔDice | Fail Rate (%) |
|---|---|---|---|---|
| Clean (baseline) | -- | 0.9145 | -- | 0.6660 |
| Gamma (γ=0.8) | Gamma | 0.9134 | −0.0011 | 0.5709 |
| Contrast (×1.2) | Contrast | 0.9139 | −0.0006 | 0.7612 |
| Contrast (×0.8) | Contrast | 0.9150 | +0.0005 | 0.5709 |
| Down-Up (×0.25) | DownUp | 0.9151 | +0.0007 | 0.3806 |
| Gamma (γ=1.2) | Gamma | 0.9172 | +0.0027 | 0.5709 |
| Noise (σ=10) | Noise | 0.9176 | +0.0031 | 0.3806 |
| Noise (σ=25) | Noise | 0.9183 | +0.0038 | 0.4757 |
| Down-Up (×0.5) | DownUp | 0.9202 | +0.0057 | 0.4757 |
| Blur (k=3) | Blur | 0.9202 | +0.0057 | 0.3806 |
| Blur (k=7) | Blur | 0.9221 | +0.0076 | 0.3806 |

*ΔDice = Dice_perturbed − Dice_clean. Failure rate = fraction of slices with Dice < 0.5. DS = downsample + upsample.*

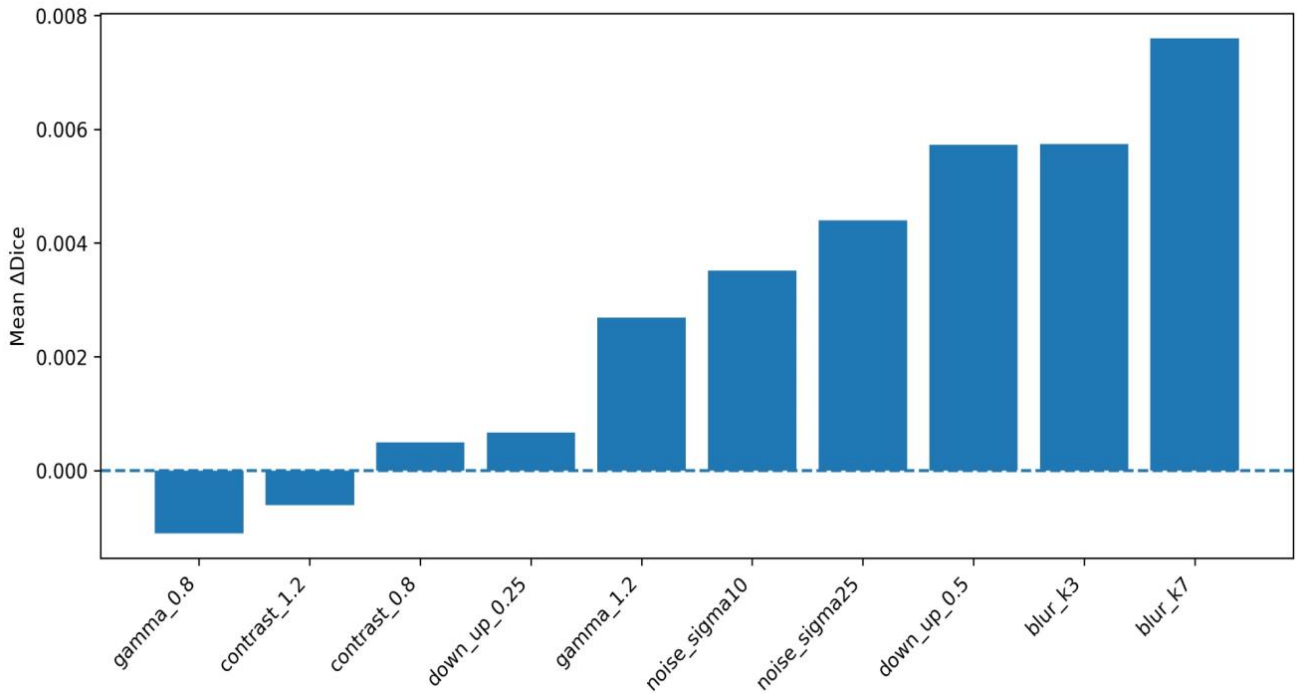

*Fig. 2. Mean change in Dice score (ΔDice = Dice_perturbed − Dice_clean) across controlled domain shift conditions (n = 1,051 slices), sorted by ascending ΔDice. Negative values (gamma_0.8, contrast_1.2) indicate modest degradation relative to the clean baseline. Positive values (blur conditions, downsampling, noise) indicate marginal improvement, likely reflecting noise suppression. All |ΔDice| < 0.008, confirming clinical stability across perturbation conditions.*

### 3.3. Failure Rate Analysis

Changes in slice-level failure status (Dice < 0.5) were evaluated using McNemar's test for paired binary outcomes, with Benjamini-Hochberg FDR correction applied independently across the 10 conditions. Compared with the clean baseline failure rate of 0.67%, no perturbation condition resulted in a statistically significant increase in failure probability (all FDR-adjusted p > 0.05). The number of discordant slice pairs (clean success to perturbed failure) remained low across all conditions, further confirming the robustness of SAM segmentation under moderate domain perturbations.

*3.4 Cumulative Reliability Profile*
The cumulative reliability profile of baseline segmentation performance is illustrated later. The CDF confirms that 81.1% of slices (852/1,051) achieved Dice ≥ 0.9, 99.3% exceeded Dice ≥ 0.5, and 99.2% exceeded Dice ≥ 0.8. These thresholds provide a quantitative basis for clinical reliability assessment prior to deployment.

## 4. Discussion

*4.1. Stability Under Moderate Perturbations*
SAM segmentation performance remained stable under all 10 evaluated perturbation conditions, with |ΔDice| < 0.01 and |rank-biserial correlation| < 0.08 across all cases. This suggests that the ViT-B image encoder produces feature representations that are relatively insensitive to moderate intensity and resolution variations, consistent with large-scale pre-training on diverse natural image data. Importantly, no condition produced a clinically meaningful degradation in mean Dice performance.

*4.2. Blur and Downsampling Effects*
The positive ΔDice observed under Gaussian blur and downsampling with upsampling likely reflects suppression of high-frequency noise artifacts in the preprocessed CT slices, which can cause subtle boundary misalignments in the SAM encoder. This does not imply improved semantic understanding but rather a smoothing benefit on noisy inputs. The effect was statistically significant but small in magnitude (|r| < 0.06) and should not be interpreted as evidence that blur universally improves SAM performance.

*4.3. SAM Integration in Health Digital Twin Pipelines*
A central motivation for this robustness audit is to evaluate SAM's suitability as a segmentation backbone within health digital twin (HDT) architectures. Digital twins for health are dynamic, patient-specific computational models that integrate multimodal clinical data streams and update their internal state continuously over time [7]. For such systems to function reliably in multi-site clinical environments, each constituent model must demonstrate predictable behavior under the full range of acquisition variability those environments produce.

In a representative HDT deployment, SAM occupies the anatomical parsing layer of a five-stage processing pipeline. Table IV summarizes the stages and the specific role SAM plays at each level.

**TABLE IV. SAM INTEGRATION WITHIN A REPRESENTATIVE HEALTH DIGITAL TWIN PIPELINE**

| Stage | Component | SAM Role |
|---|---|---|
| 1. Image Ingestion | CT scan preprocessing, HU windowing (L=50, W=400 HU), RGB conversion | Input conditioning for SAM encoder |
| 2. Foundation Segmentation | SAM ViT-B with GT-derived or automated bounding-box prompt | Organ-level binary mask generation (layer validated in this audit) |
| 3. State Update | Anatomical state repository: volumetrics, morphology, spatial anchors | Segmentation mask fed as structured anatomical state input |
| 4. Longitudinal Monitoring | Cross-timepoint change detection; threshold-based flagging | Stable Dice performance enables reliable drift detection |

| 5. Decision Support | Clinical reasoning modules: treatment planning, anomaly detection | Downstream consumer of anatomical state output |

*HDT = health digital twin; GT = ground truth; HU = Hounsfield Unit; ViT = Vision Transformer.*

At Stage 1, raw CT volumes from heterogeneous clinical scanners are ingested, windowed to a standardized Hounsfield Unit range, and converted to three-channel RGB format suitable for SAM's image encoder. At Stage 2, SAM generates organ-level binary masks using bounding-box or point prompts; this is the layer directly evaluated by the present robustness audit. At Stage 3, the resulting masks update the twin's anatomical state representation, including volumetric estimates, morphological change signals, and spatial registration anchors. Stage 4 implements longitudinal monitoring, in which the twin compares anatomical states across imaging timepoints and flags deviations that exceed a predefined clinical threshold. Stage 5 exposes the aggregated anatomical state to downstream clinical reasoning modules supporting treatment planning and anomaly detection.

The robustness properties reported in Section III directly validate Stage 2 of this architecture. A segmentation failure at Stage 2 would propagate uncorrected errors through all subsequent stages, corrupting state estimates and potentially generating false clinical alerts. The finding that $|\Delta\text{Dice}| < 0.01$ and that failure rates are non-significantly elevated across all perturbation conditions provides component-level evidence that SAM can serve as a reliable anatomical parsing backbone under the moderate inter-scanner variability characteristic of multi-site clinical environments.

Several architectural considerations inform how SAM should be integrated into production HDT systems. First, the GT-derived box-prompt protocol used here represents a controlled upper bound; practical deployments will require automated prompt generation strategies, and their robustness must be validated independently. Second, the current audit is limited to spleen segmentation; organs with greater shape variability, such as the pancreas or liver, may exhibit different sensitivity profiles and require organ-specific validation before integration. Third, longitudinal HDT applications should incorporate drift monitoring at Stage 4 to detect cumulative segmentation degradation over multiple imaging timepoints, even when individual perturbation effects appear negligible.

### 4.4. Implications for Clinical Deployment

The observed robustness under moderate domain shifts suggests that SAM may serve as a stable foundation baseline for CT segmentation tasks. However, these findings should not be interpreted as a general claim of clinical deployability. The use of GT-derived box prompts provides a controlled upper-bound estimate of encoder stability and does not reflect real-world prompting variability. Multi-institution validation, multi-organ evaluation, and comparison against supervised baselines remain necessary steps before clinical integration.

### 4.5. Cross-Model Comparison

To contextualize the spleen results reported in this study, TABLE V presents a structured comparison of SAM ViT-B performance against published specialist baselines across five medical imaging domains, conducted under the same oracle bounding-box protocol [8]. The comparison spans brain glioma segmentation in multi-parametric MRI (BraTS 2021 [10]), liver parenchyma in CT (MSD Task 03 [6]), spleen in CT (MSD Task 09 [6]), kidney tumour in CT (KiTS23 [11]), and multi-organ nuclei in H&E-stained histopathology (MoNuSeg [12]). All SAM evaluations use GT-derived bounding-box prompts following the methodology described in Section 2.2, establishing best-case zero-shot performance under idealized localization conditions consistent with this paper. Specialist reference values are drawn from nnU-Net [8] for CT organs and brain MRI, from the KiTS23 challenge benchmark [11] for kidney tumour, and from the MoNuSeg published baseline [12] for nuclei.

The results reveal a three-tier performance structure. CT parenchymal organs form the strongest tier: spleen achieves DSC 0.914 with a specialist gap of 0.049 relative to nnU-Net [8], consistent with the stability findings reported in Section 3. Kidney tumour achieves DSC 0.837 with an equivalent gap of 0.049 against the KiTS23 benchmark [11], and liver CT achieves 0.853 with a 0.090 gap and a 2.3% failure rate attributable to boundary ambiguity at the diaphragmatic dome. Brain tumour MRI drops substantially to DSC 0.515 with a 39.3% failure rate and a 0.359 specialist gap, reflecting the model's inability to distinguish active enhancing tumour from peritumoral oedema using natural-image visual priors applied to T1ce contrast [9]. H&E nuclei segmentation shows the largest gap at 0.394 Dice points and a 78.4% failure rate, driven by a fundamental mismatch between SAM's patch-based tokenization and the fine chromatin texture that defines nuclear boundaries at microscopic scale [9]. Failure rates and Dice gaps are evaluated using the framework described in Maier-Hein et al. [13].

**TABLE V. FIVE-DOMAIN CROSS-MODEL COMPARISON: SAM VIT-B VS. SPECIALIST BASELINES**

| Domain | Modality | SAM Dice (±SD) | Specialist Dice | Gap | Specialist Ref. | F<0.5 (SAM) | Dominant Failure |
|---|---|---|---|---|---|---|---|
| Spleen CT | CT | 0.914 ± 0.085 | 0.963 | 0.049 | nnU-Net [8] | 0.7% | Apical partial-volume slices |
| Kidney Tumour CT | CT | 0.837 ± 0.140 | 0.886 | 0.049 | KiTS23 [11] | 3.5% | Near-edge thin tumour slices |
| Liver CT | CT | 0.853 ± 0.137 | 0.943 | 0.090 | nnU-Net [8] | 2.3% | Diaphragmatic dome boundary |
| Brain Tumour MRI | MRI (T1ce) | 0.515 ± 0.220 | 0.874 | 0.359 | nnU-Net [8] | 39.3% | Peritumoral oedema over-extension |
| Nuclei H&E | Histopathology | 0.396 ± 0.115 | 0.790 | 0.394 | MoNuSeg [12] | 78.4% | Representational domain mismatch |

*Gap = Specialist Dice minus SAM Dice. F<0.5 = failure rate at DSC < 0.5. Specialist baselines drawn from published benchmarks; conditions are not strictly matched and gaps are approximate indicators of relative performance. All SAM evaluations conducted under GT-derived oracle bounding-box prompting consistent with Section 2.2. See [8], [9], [11], [12], [13] for full benchmark and evaluation methodology details.*

The ordering of specialist gaps (0.049 for spleen and kidney tumour, 0.090 for liver, 0.359 for brain MRI, 0.394 for nuclei) is consistent with the mechanistic account offered in Section 4.1: domains where SAM's natural-image priors translate well to the visual properties of the target exhibit small gaps, while domains where those priors are fundamentally mismatched exhibit large gaps. Notably, the spleen performance reported in this study sits within the expected range for CT parenchymal organs and represents the closest approach of zero-shot SAM to specialist-model performance across all five domains evaluated. These cross-domain findings support the interpretation that the stability results reported in Sections 3.2 through 3.4 are not exceptional artifacts of the spleen task but reflect a general property of SAM's behavior on high-contrast CT structures.

*4.6. Limitations*

This study is subject to several limitations. First, only the spleen was evaluated in the primary robustness audit; generalizability to other organs with greater shape variability (e.g., liver, pancreas) is not established within this paper, though cross-domain evidence is presented in Section 4.5. Second, GT-derived box prompts were used to isolate encoder robustness, which may overestimate practical performance with user-supplied or automated prompts. Third, perturbation levels were moderate; extreme domain shifts (severe artifact, cross-modality) were not evaluated. These limitations motivate several important directions for future investigation.

*4.7. Future Directions*

Future work should deepen the organ-level robustness analysis initiated here by evaluating additional abdominal structures with greater anatomical variability, including the pancreas and small bowel, to determine whether the stability observed for spleen extends to targets with less consistent morphology and weaker HU contrast. The cross-domain evidence presented in Section 4.5 identifies brain MRI and H&E histopathology as domains where zero-shot SAM performs poorly under oracle prompting; dedicated robustness audits for these domains under the perturbation protocol defined in Section 2.3 would characterize whether image quality variation compounds or is orthogonal to the representational failures already observed. Evaluation under automated prompting strategies represents a critical next step, as the GT-derived bounding box protocol used in this study provides a controlled upper-bound estimate of encoder robustness and does not capture real-world prompting variability. Incorporating prompt generation uncertainty into the evaluation framework will provide a more deployment-relevant assessment of model behavior.

Calibration analysis represents an additional priority: establishing whether SAM's predicted confidence scores correlate with actual segmentation quality would assess the utility of the model's own outputs for uncertainty-aware deployment. In the context of health digital twin systems, future work should examine longitudinal robustness across repeated imaging timepoints and multi-site clinical environments. Establishing stability under temporal variation and cross-institutional deployment conditions will be essential for validating segmentation models as reliable components within continuously updating patient-specific digital twin pipelines.

## 5. Conclusion

This study presented a systematic robustness audit of SAM (ViT-B) for spleen CT segmentation under 10 controlled domain shift conditions (n = 1,051 slices). Segmentation performance remained stable across all perturbations ($|\Delta Dice| < 0.01$, rank-biserial $|r| < 0.08$), and no condition produced a statistically significant increase in failure rate (McNemar, FDR-adjusted $p > 0.05$). The cumulative reliability profile confirmed that 81.1% of slices achieved Dice $\geq 0.9$ under clean conditions. The proposed standardized protocol, combining controlled perturbation, paired statistical testing, and threshold-based reliability profiling, provides a reproducible framework for evaluating foundation model stability prior to clinical deployment. From a health digital twin perspective, these findings carry direct implications for deployment readiness. Digital twins that incorporate SAM-based segmentation for organ-level anatomical modeling, whether for spleen monitoring in oncology twins, abdominal structure tracking in longitudinal population twins, or surgical planning components, can expect stable segmentation behavior across the moderate inter-scanner variability typical of multi-site clinical environments. The standardized robustness audit protocol introduced here, combining controlled domain shift perturbation, paired statistical testing, and threshold-based cumulative reliability profiling, offers a reusable validation framework that digital twin developers can apply to any segmentation component prior to integration. Establishing such component-level validation standards is a necessary step toward the governance and trustworthy deployment of health digital twins in real-world clinical and population health settings

## References

[1] A. Kirillov et al., "Segment Anything," in Proc. IEEE/CVF ICCV, 2023, pp. 4015-4026.
[2] J. Ma and B. Wang, "Segment Anything in Medical Images," Nature Commun., vol. 15, no. 1, Art. no. 654, Jan. 2024, doi: 10.1038/s41467-024-44824-z.
[3] Y. Ovadia et al., "Can You Trust Your Model's Uncertainty? Evaluating Predictive Uncertainty Under Dataset Shift," in Proc. NeurIPS, 2019.
[4] S. G. Finlayson et al., "Adversarial Attacks on Medical Machine Learning," Science, vol. 363, no. 6433, pp. 1287-1289, Mar. 2019.
[5] B. Glocker et al., "Machine Learning with Multi-Site Imaging Data: An Empirical Study on the Impact of Scanner Effects," arXiv:1910.04597, 2019.
[6] A. L. Simpson et al., "A Large Annotated Medical Image Dataset for the Development and Evaluation of Segmentation Algorithms," arXiv:1902.09063, 2019.
[7] E. Bjornsson et al., "Digital Twins for Health: A Scoping Review," npj Digital Medicine, vol. 6, no. 1, Art. no. 44, Mar. 2023, doi: 10.1038/s41746-023-00778-8.
[8] F. Isensee, P. F. Jaeger, S. A. A. Kohl, J. Petersen, and K. H. Maier-Hein, "nnU-Net: A self-configuring method for deep learning-based biomedical image segmentation," Nature Methods, vol. 18, pp. 203-211, 2021, doi: 10.1038/s41592-020-01008-z.
[9] M. A. Mazurowski, H. Dong, H. Gu, J. Yang, N. Konz, and Y. Zhang, "Segment anything model for medical image analysis: An experimental study," Medical Image Analysis, vol. 89, p. 102918, 2023, doi: 10.1016/j.media.2023.102918.
[10] B. H. Menze et al., "The multimodal brain tumor image segmentation benchmark (BRATS)," IEEE Trans. Med. Imaging, vol. 34, no. 10, pp. 1993-2024, 2015, doi: 10.1109/TMI.2014.2377694.
[11] N. Heller et al., "The KiTS23 Challenge: 500 kidney tumor cases with accompanying task hierarchy," arXiv:2307.01984, 2023.
[12] N. Kumar et al., "A multi-organ nucleus segmentation challenge," IEEE Trans. Med. Imaging, vol. 39, no. 5, pp. 1380-1391, 2019, doi: 10.1109/TMI.2019.2947628.
[13] L. Maier-Hein et al., "Metrics reloaded: Recommendations for image analysis validation," Nature Methods, vol. 21, pp. 195-212, 2024, doi: 10.1038/s41592-023-02151-z.